\documentclass[twocolumn,showpacs,preprintnumbers,amsmath,amssymb]{revtex4}

\usepackage{graphicx}
\usepackage{dcolumn}
\usepackage{bm}

\def\ii{{\mathrm{i}}}

\def\ee{{\mathrm{e}}}
\def\dd{{\mathrm{d}}}

\def\bracket#1{\langle #1 \rangle}

\def\sub#1{_{\mathrm{#1}}}
\def\up#1{^{\mathrm{#1}}}
\def\Vec#1{\mbox{\boldmath $#1$}}

\def\He4{\mbox{$^4$He}}
\def\sumint{\hbox{$\sum$}\!\!\!\!\!\!\int}

\begin{document}

\preprint{APS/123-QED}

\title{Quantum Phase Transition from Superfluid to Localized Condensates of Bose Fluid in a Confined Potential.}

\author{Michikazu Kobayashi}
\author{Makoto Tsubota}%
\affiliation{Department of Physics, Osaka City University, Sumiyoshi-Ku, Osaka 558-8585, Japan}%


\date{\today}

\begin{abstract}
We develop a model of a strongly correlated Bose fluid model in a confined potential for the purpose of analyzing the localization of Bose-Einstein condensation and the disappearance of superfluidity. This work is motivated by the recent observation of a quantum phase transition in liquid $^4$He in porous glass at high pressures. By introducing a pressure-dependent localization length of the localized condensates, we could make a new analytical criterion for the localization of the condensate. Without introducing any free parameters, the resulting critical pressure of the transition from superfluid to localization is shown to be quantitatively consistent with observations.
\end{abstract}

\maketitle

Bose-Einstein condensation (BEC) and superfluidity of liquid \He4 in random environments such as porous glass are active problems in quantum fluid research. In particular, finding out how disorder competes against long-range order of a Bose fluid has stimulated both experimental and theoretical studies.

Below 2.17 K, liquid \He4 enters both BEC and inviscid superfluid states. Superfluidity is a macroscopic quantum phenomenon like superconductivity, and understanding them has been a major goal of quantum statistical physics and low temperature physics. Various distinct features of superfluidity have been explained using the phenomenological two fluid model \cite{Tilley} in which the system consists of the inviscid superfluid and a viscous normal fluid. On the other hand, the BEC state of liquid \He4, in which a macroscopic number of particles occupies a single particle ground state and is described by a macroscopic wave function, was confirmed by neutron scattering experiments \cite{Sokol}. The inviscid superflow can be understood using this BEC wave function \cite{Huang-1}. However, the relation between the BEC and superfluidity is not necessarily understood; the one is neither necessary nor sufficient for the other. For example, in a two-dimensional Bose system, Kosterlitz and Thouless showed that superfluidity can exist even without a BEC \cite{Kosterlitz}, and this state was later observed in \He4 films \cite{Bishop}.

The random environment of a porous glass is a good system for studying the relation between BEC and superfluidity \cite{Huang-2}. In general, BEC in a random system has received considerable attention because localization effects allow some condensed particles to join the normal fluid rather than the superfluid. The phase diagram of this system has been discussed, showing a specific nonsuperfluid phase \cite{Huang-2}. Porous glass is often used as a random media in experimental studies. Vycor glass or Geltech silica is porous, having 30 to 70 \% of its volume containing wormholelike pores with diameters ranging from 25 to 100 \AA. By adjusting the pore size, one can change the strength and characteristic scale of randomness, whereas adjusting the adsorbed \He4 coverage or the applied pressure effectively changes the density of \He4 from the weakly to the strongly correlated region. By using torsional oscillators, Reppy {\it et al}. showed some remarkable features of the superfluid component and the superfluid critical temperature for changing pore sizes of porous glass or coverage \cite{Reppy, Crooker}. The superfluid density in such porous glass is smaller than that of bulk \He4 and its critical temperature decreases with the coverage. Below a certain coverage, the superfluid density can no longer exist, even near 0 K. These results show that superfluidity is broken by the random environment.

It is also important to find out how disorder affects the BEC. BEC and its elementary excitation in liquid \He4 can be observed by inelastic neutron scattering. Bulk \He4 has typical excitations consisting of phonons and rotons \cite{Tilley}. Dimeo {\it et al}., Plantevin {\it et al}. and Pearce {\it et al}. used neutron scattering and a torsional oscillator to observe the elementary excitations and the superfluid transition, respectively, of \He4 in porous glass \cite{Dimeo, Plantevin, Pearce}. Surprisingly, Plantevin {\it et al}. found that these elementary excitations existed even above the superfluid critical temperature, which is a clear evidence of BEC without superfluidity. Recently, Yamamoto {\it et al}. observed the disappearance of superfluidity of liquid \He4 confined in porous Geltech silica of pore size 25 \AA\ at pressures $P>3.5$ MPa and temperatures near 0 K without solidification \cite{Yamamoto}, suggesting a new quantum phase transition to localized BECs under the effect of a strong correlation. For Vycor glass with larger pores, this disappearance of superfluidity is suppressed by the solidification of liquid \He4

A Bose system in a random environment is also an interesting system to study theoretically. The long-range order correlation due to BEC can compete with the disorder, so that the BEC and its superfluidity are suppressed. Huang and Meng proposed a model for a three-dimensional dilute Bose gas in a random potential consisting of a small coverage of \He4 in porous glass \cite{Huang-3}. By introducing the size dependence of the random potential into their model, we could quantitatively explain the experimental results \cite{Reppy} observed by Reppy {\it et al}.; these results included the disappearance of superfluidity, even at the finite coverage and zero temperature \cite{Kobayashi}, and the presence of BEC below the critical coverage of superfluidity. Another model is the Bose-Hubbard model. This model is effective in the strongly correlated region at small densities; it includes the transfer energy, a randomly distributed on-site energy, and the on-site repulsion. Fisher {\it et al}. showed that the Bose-Hubbard model predicts the existence of the Bose glass phase, the superfluid phase, and the Mott insulating phase \cite {Fisher}. The Bose glass phase is similar to the Anderson insulating phase in metals \cite{Anderson}. In the Bose glass phase, the condensed particles are localized and thus do not contribute to superfluidity. Although some theoretical studies \cite{Krauth} examined the Bose glass phase, there is no clear experimental evidence for this phase.

This system has also been studied using a trapped alkali-atomic BEC confined in optical lattices and optical speckle beams \cite{Greiner, Lye}. The Bose-Hubbard model and the Gross-Pitaevskii model have been used to advance our understanding of this system \cite{Astrakharchik, Roth, Balabanyan}; however, some features are incompletely understood, particularly, the superfluidity and the Bose-glass phase.

In this paper, we focus on Yamamoto {\it et al}.'s \cite{Yamamoto} observation of the disappearance of superfluidity at high pressures and very low temperatures. This disappearance of superfluidity can be considered to be a pure quantum phase transition that is induced by the strong correlations and the confinement under the conditions of the experiment. However the Bose-Hubbard model is not a good model to use for these conditions for the following two reasons. First, this transition is very sensitive to pore size and it is suppressed by solidification of liquid \He4 for larger pores. We thus have to consider the pore size dependence, which is lacking in the Bose-Hubbard model. Secondly, we have to consider the liquid phase at high pressures - conditions under which the Bose-Hubbard model fails because it has only a single energy level per one site.

We start from a model of three-dimensional Bose fluid in a confined potential that depends on pore size. Then we introduce a localization length $L\sub{g}$ for localized BECs and propose a new analytical criterion for the system. We calculate the energy of the localized BECs as a function of $L\sub{g}$, and find the value $L\sub{g}=L\sub{g}\up{min}$ that minimizes the energy. The elevated pressure reduces $L\sub{g}\up{min}$ from the system size down to the order of the pore size. Then we determine the critical pressure above which $L\sub{g}\up{min}$ becomes comparable to the pore size and the superfluid component disappears. The critical pressure is found to be quantitatively consistent with the experimental value \cite{Yamamoto}, without introducing any free parameters. This result gives strong evidence that the observed disappearance of superfluidity is caused by the transition to localized BECs.

The grand canonical Hamiltonian of three-dimensional Bose-fluid in the confined potential is
\begin{align} \hat{H}-\mu\hat{N}=&\sum_{\Vec{k}}[\varepsilon(\Vec{k})-\mu]\hat{a}^\dagger(\Vec{k})\hat{a}(\Vec{k})\nonumber\\ &+\frac{1}{V}\sum_{\Vec{k}_1,\Vec{k}_2}U(\Vec{k}_1-\Vec{k}_2)\hat{a}^\dagger(\Vec{k}_1)\hat{a}(\Vec{k}_2)\nonumber\\ &+\frac{1}{2V}\sum_{\Vec{k}_1,\Vec{k}_2,\Vec{q}}\hat{a}^\dagger(\Vec{k}_1+\Vec{q})\hat{a}^\dagger(\Vec{k}_2-\Vec{q})\nonumber\\ &{}\times g_0(\Vec{q})\hat{a}(\Vec{k}_2)\hat{a}(\Vec{k}_1),\label{eq-Hamiltonian} \end{align}
where $\hat{a}(\Vec{k})$ and $\hat{a}^\dagger(\Vec{k})$ are respectively the free particle annihilation and creation operators with the discrete wave number $\Vec{k}=2\pi\Vec{n}/L$ characterized by the integer $\Vec{n}=(n_x,n_y,n_z)$. The kinetic energy is $\varepsilon(\Vec{k})=\hbar^2k^2/2m$ with $m$ the mass of one particle, $\mu$ the chemical potential, $U(\Vec{k})$ the external confined potential, $g_0(\Vec{k})$ the interparticle interaction, and $V=L^3$ the volume of the cubic system with size $L$. We assume that number $N_0$ of particles are condensed at the lowest wave number $\Vec{k}=0$ and thus replace the operators of Bose particle $\hat{a}(\Vec{k}=0)$ and $\hat{a}^\dagger(\Vec{k}=0)$ with $\sqrt{N_0}$. The second and third terms of the right-hand side of Eq. (\ref{eq-Hamiltonian}) represent the interaction between a particle and the confined potential and the interparticle scattering, respectively. Here we treat $U(\Vec{k})$ using the second-order perturbation theory and $g_0(\Vec{k})$ by the ring-approximation, which is one of the simplest method for strongly correlated system of high density. For these calculations, we introduce the Green function
\begin{equation} \hat{G}(k_0,\Vec{k})=\left[\begin{array}{cc} G_{11}(k_0,\Vec{k}) & G_{12}(k_0,\Vec{k})\\ G_{21}(k_0,\Vec{k}) & G_{22}(k_0,\Vec{k})\end{array}\right],\label{eq-Green-function} \end{equation}
with $\Vec{k}\neq 0$. Calculation of $\hat{G}$ starts from the non-perturbative Green function
\begin{align} \hat{G}^0(k_0,\Vec{k})=&\left[\begin{array}{cc} G_{11}^0(k_0,\Vec{k}) & G_{12}^0(k_0,\Vec{k})\\ G_{21}^0(k_0,\Vec{k}) & G_{22}^0(k_0,\Vec{k})\end{array}\right]\nonumber\\ =&\left[\begin{array}{cc} \hbar/\xi(k_0,\Vec{k}) & 0\\ 0 & \hbar/\xi(-k_0,-\Vec{k})\end{array}\right],\label{eq-Green-0-function} \end{align}
with $\xi(k_0,\Vec{k})=\hbar k_0-\varepsilon(\Vec{k})+\mu+\ii 0$. The effective interparticle interaction $g(q_0,\Vec{q})$ with the ring approximation is given by
\begin{equation} g(q_0,\Vec{q})=\frac{g_0(\Vec{q})}{\displaystyle 1-\frac{N_0}{\hbar V}g_0(\Vec{q})G_{11}^0(q_0,\Vec{q})}.\label{eq-effective-interaction} \end{equation}
The Green function $\hat{G}\up{I}(k_0,\Vec{k})$ that includes this effective interparticle interaction can be obtained from the following Dyson equation:
\begin{equation} \hat{G}\up{I}(k_0,\Vec{k})=\hat{G}^0(k_0,\Vec{k})+\hat{G}^0(k_0,\Vec{k})\hat{\Sigma}(k_0,\Vec{k})\hat{G}\up{I}(k_0,\Vec{k}),\label{eq-Dyson} \end{equation}
where $\hat{\Sigma}(k_0,\Vec{k})$ is the self energy defined as
\begin{subequations}
\begin{align} \hat{\Sigma}(k_0,\Vec{k})=&\left[\begin{array}{cc} \Sigma_{11}(k_0,\Vec{k}) & \Sigma_{12}(k_0,\Vec{k})\\ \Sigma_{21}(k_0,\Vec{k}) & \Sigma_{22}(k_0,\Vec{k})\end{array}\right],\label{eq-self-energy-1}\\ \Sigma_{11}(k_0,\Vec{k})=&\Sigma_{22}(-k_0,-\Vec{k})\nonumber\\ =&\frac{N_0}{\hbar V}[g_0(0)+g_0(\Vec{k})]\nonumber\\ &{}-\frac{1}{\hbar V}\frac{1}{2\pi\ii}\sumint_{\Vec{q}\neq 0, \Vec{k}}\dd q_0\nonumber\\ &\quad\times G_{11}^0(q_0,\Vec{q})g(k_0-q_0,\Vec{k}-\Vec{q}),\label{eq-self-energy-2}\\ \Sigma_{12}(k_0,\Vec{k})=&\Sigma_{21}(k_0,\Vec{k})=\frac{N_0}{\hbar V}g_0(\Vec{k}).\label{eq-self-energy-3} \end{align}
\end{subequations}
Finally, we can obtain the Green function $G_{11}(k_0,\Vec{k})$ with the second-order perturbation of $U(\Vec{k})$ as follows:
\begin{align} G_{11}(k_0,\Vec{k})=&G_{11}\up{R1}(k_0,\Vec{k})+G_{11}\up{R2}(k_0,\Vec{k}),\nonumber\\ G_{11}\up{R1}(k_0,\Vec{k})=&-\frac{(2\pi\ii)\delta(k_0)N_0}{\hbar^2V^2}[G_{11}\up{I}(k_0,\Vec{k})+G_{12}\up{I}(k_0,\Vec{k})]\nonumber\\ &\quad\times[G_{11}\up{I}(k_0,\Vec{k})+G_{21}\up{I}(k_0,\Vec{k})]|U(\Vec{k})|^2,\nonumber\\ G_{11}\up{R2}(k_0,\Vec{k})=&\frac{1}{\hbar^2V^2}\sum_{\Vec{q}\neq 0}[G_{11}\up{I}(k_0,\Vec{k})G_{11}\up{I}(k_0,\Vec{q})G_{11}\up{I}(k_0,\Vec{k})\nonumber\\ &+G_{11}\up{I}(k_0,\Vec{k})G_{12}\up{I}(k_0,\Vec{q})G_{21}\up{I}(k_0,\Vec{k})\nonumber\\ &+G_{12}\up{I}(k_0,\Vec{k})G_{21}\up{I}(k_0,\Vec{q})G_{11}\up{I}(k_0,\Vec{k})\nonumber\\ &+G_{12}\up{I}(k_0,\Vec{k})G_{22}\up{I}(k_0,\Vec{q})G_{21}\up{I}(k_0,\Vec{k})]\nonumber\\ &\quad\times |U(\Vec{k}-\Vec{q})|^2.\label{eq-potential-perturbation} \end{align}
The chemical potential is given by the Hugenholtz-Pines theorem \cite{Hugenholtz}:
\begin{align} \mu=&\frac{N}{V}U(0)+\frac{N_0}{V}g_0(0)\nonumber\\ &-\frac{1}{V}\frac{1}{2\pi\ii}\sumint_{\Vec{q}\neq 0}\dd q_0\: G_{11}^0(q_0,\Vec{q})g(-q_0,-\Vec{q}).\label{eq-Hugenholtz-Pines} \end{align}
The number of condensate particles and the total energy are
\begin{align} N_0=&N+\frac{1}{2\pi\ii}\sumint_{\Vec{k}\neq 0}\dd k_0\:G_{11}(k_0,\Vec{k})\ee^{+\ii k_0 0},\label{eq-total-particle}\\ E=&\frac{1}{2}\mu N\nonumber\\ &-\frac{1}{4\pi\ii}\sumint_{\Vec{k}\neq 0}\dd k_0\:[\ii\hbar k_0+\varepsilon(\Vec{k})+\mu]G_{11}(k_0,\Vec{k})\ee^{+\ii k_0 0}.\label{eq-total-energy} \end{align}
Here $N$ is the total number of particles.

The calculation of superfluidity is based on the two fluid model in which the number $N$ of particles equals the number in the normal fluid component $N\sub{n}$ and the superfluid component $N\sub{s}$. The superfluid component $N\sub{s}$ can be calculated using linear response theory \cite{Hohenberg}. Because of its viscosity, only the normal fluid responds to a small, applied velocity field. Thus, the normal fluid component can be defined by the response of the momentum density to the external velocity field. After some tedious calculations, the linear response theory gives the final form of $N\sub{n}$ and $N\sub{s}$ as
\begin{align} N\sub{n}=&\frac{\hbar}{3m}\frac{1}{2\pi\ii}\sumint_{\Vec{k}\neq 0}\dd k_0\:k^2\nonumber\\ &\times[G_{11}(k_0,\Vec{k})^2-G_{12}(k_0,\Vec{k})G_{21}(k_0,\Vec{k})]\ee^{+\ii k_0 0},\nonumber\\ N\sub{s}=&N-N\sub{n}\label{eq-linear-response}. \end{align}
In the derivation, we assumed the rotational invariance. This derivation of $N\sub{n}$ and $N\sub{s}$ is consistent with the Khalatnikov's method based on Galilean invariance \cite{Khalatnikov}.

Introducing a new criterion, we consider localized BECs. We divide the total cubic system of the volume $V=L^3$ into cubic subsystems of the same shape except the volume $V\sub{g}=L\sub{g}^3$. Then we assume that every cubic subsystem has the same localized BEC and the number of particles $N\sub{g}=NV\sub{g}/V$. The energy $E\sub{g}$ of one subsystem can be obtained by replacing the volume $V$ and the number of particles $N$ in the above equation with $V\sub{g}$ and $N\sub{g}$. Because the number of the localized BECs is $V/V\sub{g}$, the energy of the total system should be $E\sub{g}V/V\sub{g}$. The ideal or weakly interacting Bose gas has $E\sub{g}$ that is proportional to $V\sub{g}$, whereas in the strongly correlated Bose fluid, $E\sub{g}$ is no longer a simple linear function of $V\sub{g}$; thus, $E\sub{g}/V\sub{g}$ depends on $V\sub{g}$. Calculating $E\sub{g}/V\sub{g}$ as a function of $V\sub{g}$, we determine the value $V\sub{g}=V\sub{g}\up{min}$ that minimizes the total energy $E=E\sub{g}V/V\sub{g}$. Then, the localization of BEC can be defined by the following criterion.
\begin{itemize}
\item If $V\sub{g}\up{min}$ exceeds $V$, the system is a non-localized BEC state.
\item When $V\sub{g}\up{min}$ is reduced to the pore size, we assume that the BEC is localized within the localization length $L\sub{g}\up{min}=\sqrt[3]{V\sub{g}\up{min}}$
\end{itemize}

To apply this criterion to the experiment by Yamamoto {\it et al}. \cite{Yamamoto}, we numerically calculated the condensate particle $N_0$, superfluid component $N\sub{s}$ and the localization volume $V\sub{g}\up{min}$ as a function of the pressure $P=-\partial E/\partial V$ by changing the number of particles $N$, and then compared our results with Yamamoto {\it et al}.'s experiment \cite{Yamamoto}. To make a quantitative comparison, we used the following numerical parameters. The mass of a \He4 atom is $m\simeq 6.6\times 10^{-27}\:\mathrm{kg}$. For the bare interparticle interaction, we used the simple gaussian interaction $g_0(\Vec{k})=\nu_0(\sigma\sqrt{2\pi})^3\exp[-k^2\sigma^2/2]$, where parameters $\nu_0\simeq 2.1\times 10^{-17}\:\mathrm{J}$ and $\sigma\simeq 2.5\times 10^{-11}\:\mathrm{m}$ were determined using a comparison to the effective interparticle potential \cite{Aziz} proposed by Aziz {\it et al}. The total volume of Geltech silica used in Yamamoto {\it et al}.'s experiment was $V\simeq 5.9\times 10^{-8}\:\mathrm{m}^3$. For the external confined potential $U(\Vec{k})$, we used the random distribution
\begin{equation} \bracket{U^\ast(\Vec{k})U(\Vec{k})}=U_0^2\exp\Big[-\frac{k^2r\sub{p}^2}{4\pi^2}\Big].\label{eq-random-potential} \end{equation}
We then calculate physical values for many ensemble of $U(\Vec{k})$ and took an arithmetic average. In Eq. (\ref{eq-random-potential}), $r\sub{p}\simeq 2.5\times 10^{-9}\:\mathrm{m}$ is the pore size of Geltech silica. We chose the characteristic strength of the potential $U_0\simeq 2.0\times 10^{-32}\:\mathrm{J}$ to ensure the critical adsorbed liquid \He4 coverage $n\sub{c}\simeq 20\:\mu\mathrm{mol}/\mathrm{m^2}$ per pore surface area below which superfluidity also disappears, as we did in our previous study \cite{Kobayashi}.

\begin{figure}[t] \begin{center} \begin{minipage}{0.49\linewidth} \begin{center} \includegraphics[width=0.99\linewidth]{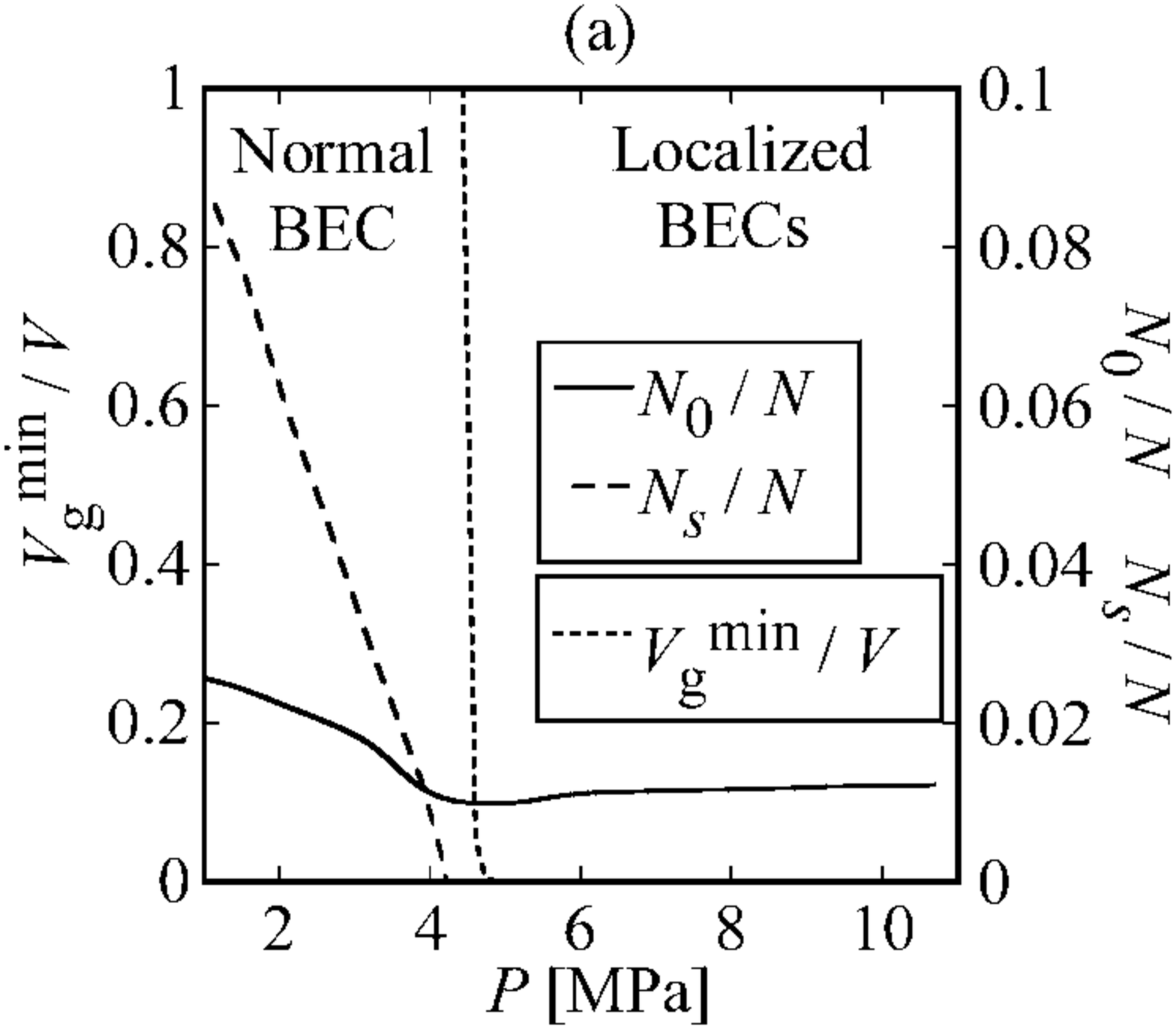} \end{center} \end{minipage} \begin{minipage}{0.49\linewidth} \begin{center} \includegraphics[width=0.99\linewidth]{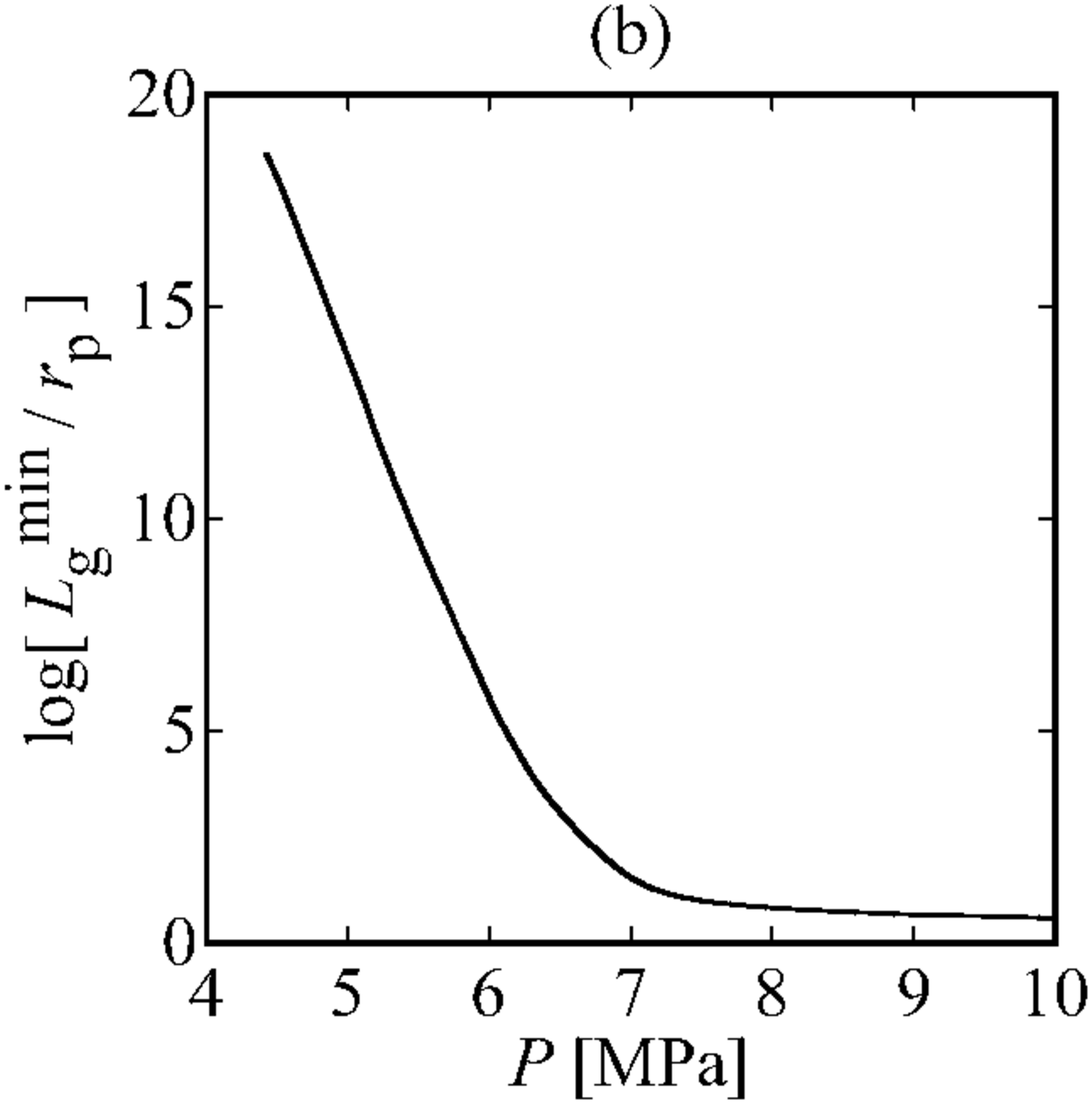} \end{center} \end{minipage} \end{center}
\caption{\label{fig-local-BEC} Calculation results. (a) Dependence of the condensate particle number $N_0$, the superfluid component $N\sub{s}$, and the volume $V\sub{g}\up{min}$ of the localized BECs on the pressure. (b) Log-plot of the dependence of $L\sub{g}\up{min}$. on the pressure} \end{figure}

Because all parameters were fixed, we can quantitatively compare calculations to experiment. Figure \ref{fig-local-BEC}(a) shows the pressure dependence of $N_0$, $N\sub{s}$, and $V\sub{g}\up{min}$. At the pressure $P\sim 4.2\:\mathrm{MPa}$, the superfluid component $N\sub{s}$ disappears and $V\sub{g}\up{min}$ decreases dramatically. We therefore define this pressure as the critical pressure $P\sub{c}$. The condensate particle number $N_0$ also decreases just below $P\sub{c}$. This number recovers above $P\sub{c}$, although the condensates are localized. The dependence of $L\sub{g}\up{min}$ in Fig. \ref{fig-local-BEC} (b) shows that the localization length $L\sub{g}\up{min}$ reduces to about the pore size $r\sub{p}$. Moreover, the critical pressure $P\sub{c}\sim 4.2\:\mathrm{MPa}$ is nearly equal to the experimental value of $P\sub{c}\sim 3.5\:\mathrm{MPa}$. Therefore, we conclude that the experimental disappearance of superfluidity at high pressures is caused by the transition from a normal BEC to localized BECs. We made a similar calculation for the case of porous Vycor glass, which has larger pores than Geltech silica. Using the numerical parameters $r\sub{p}\simeq 7.0\times 10^{-9}\:\mathrm{m}$ and $U_0\simeq 8.9\times 10^{-33}\:\mathrm{J}$, we obtained the much larger critical pressure of $P\sub{c}\sim 9\:\mathrm{MPa}$, which is too high for a BEC to be localized against solidification of liquid \He4. This is also consistent with experimental results.

In conclusion, we investigated the localization of BEC and the disappearance of superfluidity in a strongly correlated Bose fluid at high pressures. In the model, the BEC is assumed to be localized in the pores when the pressure-dependent localization length equals the pore size. Without introducing any free parameters, we showed that the critical pressure above which BEC localizes is quantitatively consistent with the experimental result of liquid \He4 in porous Geltech silica. Our criterion can be generally applicable to many systems with localization or quasi-condensation. For example, this formulation may be closely connected with the renormalization group analysis of the Kosterlitz-Thouless transition \cite{Kosterlitz}.

We acknowledge Keiya Shirahama for many useful discussions.

\end{document}